\documentclass[11pt,oneside,nofootinbib]{article}
\usepackage{a4wide}
\usepackage{mathrsfs}
\usepackage{latexsym,bm}
\usepackage{graphicx}
\usepackage{indentfirst}
\usepackage{slashed}
\usepackage{amsmath}
\usepackage{amssymb}
\usepackage{color}
\usepackage{hyperref}
\usepackage{epsfig}
\usepackage[titletoc]{appendix}
\usepackage{multirow}%
\usepackage{rotating}
\usepackage{cite}
\usepackage[normalem]{ulem} 
\usepackage[top=2.9cm,bottom=2.5cm,left=2.8cm,right=3cm]{geometry}
\usepackage{tabularx}

\newcommand{\email}[1]{\footnote{{\em } \texttt{#1}}}

\newcommand{\jpsi}{J/\psi}

\newcommand{\dvv}{D^{*}}
\newcommand{\dvvb}{\bar{D}^{*}}
\newcommand{\db}{\bar{D}}
\newcommand{\xic}{\Xi_c}
\newcommand{\xicp}{{\Xi'}_c}

\newcommand{\pcs}{P_{cs}(4459)}

\newcommand{\ra}{\rangle}

\newcommand{\nn}{\nonumber}

\newcommand{\jo}[1]{\textcolor{black}{{{#1}}}}

\begin{document}

\thispagestyle{empty}
\title{
\Large \bf Insights into the nature of the $P_{cs}(4459)$ }
\author{\small  Meng-Lin Du$^{a}$\email{du.menglin@ific.uv.es}\,, Zhi-Hui Guo$^{b}$\email{zhguo@seu.edu.cn}\,, J.~A.~Oller$^{c}$\email{oller@um.es}  \\[0.5em]
{\small \it ${}^a$ Instituto de F\'isica Corpuscular (centro mixto CSIC-UV), } \\
{\small\it Institutos de Investigaci\'on de Paterna, Apartado 22085, 46071, Valencia, Spain}\\[0.2em]
{ \small\it ${}^b$ School of Physics, Southeast University, Nanjing 211189, China }\\[0.2em]
{\small \it ${}^c$ Departamento de F\'{\i}sica. Universidad de Murcia. E-30071 Murcia. Spain}
}
\date{}

%

\maketitle
\begin{abstract}
  We study the nature of the recently observed $P_{cs}(4459)$ by the LHCb collaboration by employing three methods based on the elastic effective-range expansion and the resulting size of the effective-range, the saturation of the compositeness relation and width of the resonance, and a direct fit to data involving the channels $J/\psi\Lambda$, $\Xi'_c\bar{D}$ and $\Xi_c\bar{D}^*$. We have also considered the addition of a CDD pole but this scenario can be discarded. Our different analyses clearly indicate the molecular nature of  the $P_{cs}(4459)$ with a clear  $\Xi_c\bar{D}^*$ dominant component. In relation with heavy-quark-spin symmetry our results also favor the actual existence of two resonances with $J=1/2$ (the lighter one) and $3/2$ (the heavier one) in the energy region of the $P_{cs}(4459)$. In the scenario of two-resonance \jo{for the} $P_{cs}(4459)$, the inclusion of the $\Xi_c'\bar{D}$ channel is required for the their mass \jo{splitting} and \jo{it} allows one to \jo{determine} the spin structures of the two resonances. 
\end{abstract}

\section{Introduction}

The study of exotic hadrons  beyond the conventional quark-antiquark ($q\bar{q}$) configurations for mesons and the three-quark ($3q$) ones for baryons has been an interesting topic in strong interactions since long time ago, e.g. one could mention the discussions on the lightest scalar resonances $f_0(500)$, $f_0(980)$, $a_0(980)$ and $\kappa(900)$ \cite{Jaffe:1976ig,Jaffe:1976ih,Weinstein:1983gd,npa620,Oller:2003vf}, the case of the $\Lambda(1405)$ resonances \cite{Dalitz:1959dn,Oller:2000fj,Mai:2020ltx}, the Roper \cite{Burkert:2017djo}, etc. The interest on this topic has received a great boost  since  the discoveries in 2003 of the $D_{s0}^*(2317)$ \cite{BaBar:2003oey} and the $\chi_{c1}(3872)$ \cite{Belle:2003nnu}, aka $X(3872)$,  whose properties cannot be accommodated within the conventional quark model expectations. As a result a new era has developed in hadron spectroscopy based on the detection of numerous multiquark states with heavy quarks, particularly the charm quark, and more recently the bottom quark as well.

An important outcome in this fruitful field of research was the discovery by the LHCb in 2015 of the charmonium pentaquarks $P_c(4450)$ and $P_c(4380)$ decaying into $J/\psi p$ \cite{LHCb:2015yax}. With higher statistics in 2019 \cite{LHCb:2019kea} the same collaboration in the same decay process determined the existence
of the pentaquarks $P_c(4457)$, $P_c(4440)$ (instead of just one resonance $P_c(4450)$) and the lighter and narrow $P_c(4312)$. In Refs.~\cite{Du:2021fmf,Du:2019pij} a through analysis of the experimental data employing coupled-channel dynamics, heavy-quark-spin symmetry (HQSS) and one-pion exchange provided a clear hint for the existence  of a narrow $P_c(4380)$.

Very recently the LHCb collaboration \cite{Aaij:2020gdg} has reported the first evidence of a charmonium pentaquark resonance with strangeness, the so called $P_{cs}(4459)$ appearing as a peak in the $J/\psi \Lambda$ event distribution from an amplitude analysis of the $\Xi_b^-\to J/\psi\Lambda K^-$ decay. Ref.~\cite{Aaij:2020gdg} also concludes that this structure is consistent with two resonances. Indeed strange pentaquark resonances were already predicted theoretically in several references \cite{Wu:2010jy,Chen:2016ryt,Santopinto:2016pkp,Shen:2019evi,Xiao:2019gjd,Wang:2019nvm,Wang:2015wsa,Anisovich:2015zqa}, being also suggested to look for them in the decay process $\Xi_b^-\to J/\psi\Lambda K^-$ by the Refs.~\cite{Chen:2015sxa,Santopinto:2016pkp}. 

The discovery of the $P_{cs}(4459)$ by the LHCb collaboration has prompted numerous studies of this resonance \cite{Yang:2021pio,Peng:2020hql,Chen:2020uif,Chen:2020kco,Zhu:2021lhd,Lu:2021irg}. These studies typically favor a molecular-type picture  with a dominant component of the channel   $\Xi_c\bar{D}^*$, whose threshold lies only about 19~MeV above the nominal mass of the $P_{cs}(4459)$. However, there is no such a general consensus regarding the spin of the resonance between the two values $J=1/2$ and $3/2$ and, indeed, some studies point out towards the existence of two resonances around 4.45~GeV in the $J/\psi \Lambda$ event distributions. 

In this work we intent to scrutinize the nature of the $P_{cs}(4459)$ \jo{without ascribing to a specific dynamical model}, 
by making use of general arguments based on unitarity and analyticity in $S$-matrix theory.
We exploit three general methods, \jo{recently applied in a similar way by two of us in the study of the $X(6900)$ \cite{Guo:2020pvt}}, all of which come to the conclusion of the molecular nature of the $P_{cs}(4459)$ with a clearly dominant $\Xi_c\bar{D}^*$ component.

The first method is based on the fact that a  CDD pole \cite{castillejo} in the near-threshold region gives rise to a strong impact \cite{Kang:2016ezb} in the values of the scattering length $a$ and  effective range $r$. Particularly relevant is to focus on $r$ because a priori its absolute value is given by the typical range of strong interactions around 1~fm \cite{Bethe:1949yr}, while the presence of a nearby CDD pole would  typically imply much larger values \cite{Kang:2016ezb}. This signature could then be taken as favoring the presence in the state of interest of a large bare component, or of large components from other channels, to which the CDD pole would be associated \cite{Morgan:1992ge}. However, we find here that $r$ has a natural value around $-1$~fm  by applying the elastic effective-range-expansion (ERE) study to $\Xi_c\bar{D}^*$, suggesting the  $\Xi_c\bar{D}^*$ molecular nature of this resonance. We also observe that when this method is applied to the $\Xi_c'\bar{D}$ channel (whose threshold is around 13~MeV below the resonance mass) the resulting $r$ has instead a much larger modulus indicating that the $P_{cs}(4459)$ is not dynamically generated by this channel. 

The second  method is a coupled-channel generalization that relies on the saturation of the width $\Gamma$  and compositeness $X$ of the resonance in terms of the $J/\psi\Lambda$ and $\Xi_c\bar{D}^*$($\Xi_c'\bar{D})$ channels. The latter concept refers to the weight of the baryon-meson components in the $P_{cs}(4459)$. For its calculation we follow Ref.~\cite{Guo:2015daa} that provides a probabilistic interpretation for the compositeness of a resonance. The optimal way to calculate the compositeness  is still under debate \cite{Hyodo:2011qc,Aceti:2012dd,Sekihara:2014kya,Guo:2015daa,Matuschek:2020gqe,Matuschek:2020gqe}. By following this method we have at our disposal two equations,  though the outcome depends on the value assumed for the total compositeness $X$, while $\Gamma$ is taken from the LHCb collaboration Ref.~\cite{Aaij:2020gdg}. Nonetheless, given a value of $X$ the method provides us with the partial decay widths and the partial compositeness coefficients for each channel.

The third method allows one to also include coupled channels and avoid having to take for granted a value for $X$. We proceed with  a direct fit to the $J/\psi \Lambda $ event distribution around the $P_{cs}(4459)$ resonance energy region explicitly taking into account the three coupled channels $J/\psi \Lambda$, $\Xi_c'\bar{D}$ and $\Xi_c\bar{D}^*$, together with constraints from HQSS. The lightest channel accounts in an effective way of those channels whose thresholds are relatively far from the resonance mass, in particular  this is the case for the $\eta_c\Lambda$ which is degenerate to the $J/\psi\Lambda$ in the heavy-quark limit. Our analysis also favors the presence of two resonances (with $J=\frac12$ and $J=\frac32$ respectively) in the peak of the $J/\psi \Lambda$ event distribution associated to the $P_{cs}(4459)$. Since the $S$-wave $\Xi_c'\bar{D}$ only couples to $\Xi_c\bar{D}^*$ in  $J=\frac12$, it could affect the dynamics of this channel and leave significant impact on the line shape.
As a result, it leads to the mass split of the two resonances and allows one to 
determine their spins.
We stress that the HQSS breaking of the potential, which is of order $\mathcal{O}(\frac{\Lambda_\text{QCD}}{m_c})$, cannot account for the mass \jo{splitting} of the two resonances, \jo{cf.} Sec.~\ref{sec.210803.2}.
In other words, \jo{the inclusion of the $\Xi_c'\bar{D}$ channel}  provides an extra mechanism \jo{which is} responsible
for the mass \jo{splitting} of the two resonances \jo{while keeping the HQSS constraints on the interacting potential}.  
In addition, we have also allowed the presence of a CDD pole in the $\Xi_c\bar{D}^*$ or $\Xi_c'\bar{D}$  channel but no acceptable fits result. 

We find that the methods 1 and 3 agree in the basic fact that the $P_{cs}(4459)$ is mainly a $\Xi_c\bar{D}^*$ molecular resonance, while the  methods 2 and 3 provide also very compatible values for the couplings of the resonance to the different channels when the same value for $X$ is used in the second method. 

The contents of the manuscript are structured as follows. After this introduction we describe and apply the methods 1, 2 and 3 to the study the $P_{cs}(4459)$ in the Secs.~\ref{sec.ere}, \ref{sec.threech} and \ref{sec.fits}, respectively. In the latter section we also compare between several types of fits with different number of channels and partial waves involved, as well as with different degree of imposition of the HQSS constraints. The conclusions are gathered in Sec.~\ref{sec.conc}.

\section{Effective range expansion for the elastic scattering }\label{sec.ere}

First we rely on the ERE method to perform an exploratory study of the $\pcs$ near the $\xic\dvvb$ and $\xicp\db$ thresholds. The elastic $S$-wave ERE formula reads 
\begin{eqnarray}\label{eq.til}
T(E)=\frac{1}{-\frac{1}{a}+\frac{1}{2}r\,k^2-i\,k}\,,
\end{eqnarray}
where $a$ and $r$ are the scattering length and the effective range in order, and the non-relativistic relation between the three-momentum $k$ and the energy $E$ in the center of mass (CM) frame is 
\begin{eqnarray}
k= \sqrt{2\mu_m(E-m_{\rm th})}\,,
\end{eqnarray}
being $\mu_m=m_1m_2/(m_1+m_2)$ the reduced mass and $m_{\rm th}=m_1+m_2$ the threshold of the two states in the scattering. 
Above the threshold, the unitarity relation 
\begin{equation}
 {\rm Im}\,T(E)^{-1} = - k  \,,\quad (E > m_{\rm th})\,,
\end{equation}
is automatically satisfied by the amplitude in Eq.~\eqref{eq.til}. 

There are two caveats when applying the ERE formalism. First the left hand cuts originated from the particle exchanges in the crossed channels are neglected. This assumption is consistent with the pionless effective filed theory description, which turns out to be able to explain the various pentaquark candidates~\cite{Liu:2019tjn,Peng:2021hkr}. Nevertheless, further studies on the inclusion of the light-meson exchanges reveal possibly noticeable effects on the assignments of the spin for the different pentaquark states~\cite{PavonValderrama:2019nbk,Liu:2019zvb,Du:2019pij,Du:2021fmf}. The other caveat is about the possible underlying CDD pole near the threshold. It is rather challenging to infer the existence of the near-threshold CDD pole a priori. In Refs.~\cite{Guo:2016wpy,Kang:2016ezb}, a practicable criteria is provided to  discern 
the appearance of a near-threshold CDD pole based on 
the large magnitude of the effective range $r$.
This is because the effective range is found to scale as 
\begin{eqnarray}\label{eq.rcdd}
r \propto \frac{1}{(M_\text{CDD}-m_{\rm th})^2}\,,
\end{eqnarray}
when the CDD position in energy, $M_\text{CDD}$, approaches to the threshold.
To be more specific, if the resulting magnitude of $r$ in the ERE study is large,
as compared to the typical range around 1~fm  of strong interactions,
it would be a clear hint for the existence of a CDD pole near threshold,
indicating as well  that the ERE expansion is not suitable to handle such systems.
On the contrary, if the resulting $|r|$ has a value less than $1\sim 2$~fm,
a near-threshold CDD pole is not necessary and  the ERE would not be spoiled by its presence, and it could be
used to describe such scattering processes.~\footnote{Note that a large value of $|r|$ can also be caused by a near coupled-channel effect. 
 \jo{However, this effect could also be related to the presence of a CDD pole in the elastic scattering of the open channel.}}

Regarding the previous discussion, by identifying the observed pentaquark candidate $\pcs$ as the resonance pole in the {\it elastic}  scattering amplitude of $\xic\dvvb$ or $\xicp\db$, one can then fix in  every case  the corresponding scattering length and effective range via the mass and width of the $\pcs$. The size obtained for the effective range would be an indication of whether the resonance can be considered as ``molecular'' in terms of the channel considered or it better corresponds to a bare state (preexisting resonance). 
 The resonance pole lies in the second Riemann sheet (RS) in the elastic scattering case and the amplitude $T_{II}(E)$ in this RS has the form
\begin{eqnarray}\label{eq.tiil}
 T_{II}(E) = \frac{1}{-\frac{1}{a}+\frac{1}{2}r\,k^2\,+i\,k}\,.
\end{eqnarray}
In this convention, the imaginary part of $k$ is understood to be positively  taken, namely ${\rm Im}k>0$, in Eqs.~\eqref{eq.til} and \eqref{eq.tiil}. The solution of $T_{II}(E_R)^{-1}=0$ gives the resonance pole position $E_R=M_R-i\Gamma_R/2$, being $M_R$ and $\Gamma_R$ the resonance mass and width, respectively. Alternatively, one could also determine $a$ and $r$ in terms of the resonance pole position $E_R$ and their explicit relations turn out to be~\cite{Guo:2016wpy} 
\begin{eqnarray}\label{eq.ar0}
a=-\frac{2k_i}{|k_r|^2+|k_i|^2}\,,\quad r=-\frac{1}{k_i} \,,
\end{eqnarray}
where $k_r$ and $k_i$ stand for the real and imaginary parts of the three-momentum $k_R$ evaluated at the pole, namely, $k_R=\sqrt{2\mu_m(E_R-m_{\rm th})}$. For clarity, let us point out again that the convention $k_i>0$ is taken.

The ERE scattering amplitude is completely fixed after $a$ and $r$ are known, and with it an important quantity, i.e. the residue, which can then be inferred by taking the Laurent expansion of $T_{II}(E(k))$ around the pole position   
\begin{eqnarray}
T_{II}(k)=\frac{-k_i/k_r}{k-k_R}+ {\cal O}[(k-k_R)^0]\,.
\end{eqnarray} 
In our previous studies~\cite{Guo:2015daa,Kang:2016ezb}, this residue $-k_i/k_r$ has been demonstrated to be equal to the compositeness coefficient $X$ \cite{Guo:2015daa}, which corresponds to the probability of the two-body component in the resonance, i.e.,  
\begin{eqnarray}\label{eq.x}
 X= -\frac{k_i}{k_r}\,.
\end{eqnarray} 
As proven in Ref.~\cite{Kang:2016ezb} when $M_R-m_{\rm th}>0$  it follows that
the compositeness $X=-k_i/k_r$ lies within the range $[0,1]$, 
a result that is also consistent with the working condition of the probabilistic interpretation revealed in Ref.~\cite{Guo:2015daa}. It should be stressed that the expression of the compositeness coefficient $X$ in Eq.~\eqref{eq.x} is only valid in the context of the single-channel ERE when there is a resonance pole in the amplitude. General discussions on the compositeness relations for resonances will be given in the next section.

The mass and width of the $P_{cs}(4459)$ from the single-resonance fit of the LHCb collaboration \cite{Aaij:2020gdg} read 
\begin{eqnarray} 
\label{lhcb.mods}
 M_R=4458.8\pm 2.9^{+4.7}_{-1.1}~\text{MeV}\,,\qquad \Gamma_R=17.3\pm 6.5^{+8.0}_{-5.7}~\text{MeV}\,, 
\end{eqnarray}
where the first errors are statistics and the second ones are systematic.
We will take a conservative estimation and combine the two types of uncertainties by adding in quadrature the statistic error and the largest of the two systematic errors. The explicit values of the mass and width used in our analyses are given in the second and third columns of Table~\ref{tab.ar}, respectively, together with the thresholds of $\xicp\db$ and $\xic\dvvb$ in the next column.

\begin{table}[htbp]
\centering
\begin{scriptsize}
\begin{tabular}{ c c c c c c c}
\hline\hline
 Resonance & Mass   & Width & Threshold  & $a$     & $r$  & $X$    \\
           & (MeV)  & (MeV) &  (MeV)    & (fm)    &  (fm)   
\\ \hline \\
$P_{cs}$ & $4458.8\pm 5.5$ & $17.3\pm 10.3$ & ${\Xi'}_c D$~(4446.0)  & $-0.63\pm 0.38$ &  $-3.68\pm 2.11$ & $0.31\pm 0.19$  \\ \hline \\
$P_{cs}$ & $4458.8\pm 5.5$ & $17.3\pm 10.3$ & $\Xi_c D^*$~(4478.0)  & $-1.79\pm 0.23$ & $-0.94\pm 0.13$ & $--$
\\  \hline
\end{tabular}
\end{scriptsize}
\caption{Results for the scattering length, effective range and the compositeness coefficient from the elastic ERE studies given in the last three columns in this order. The channel involved in the scattering is indicated in the fourth column. \label{tab.ar}} 
\end{table}

As a tentative study,  let us now focus on the two elastic scattering cases to address the $\pcs$, i.e. those stemming from taking separately 
the $\xicp\db$ and $\xic\dvvb$, whose thresholds lie near the $\pcs$ peak observed by the LHCb collaboration~\cite{Aaij:2020gdg}. Therefore, we separately consider in the ERE analysis that the $\pcs$ corresponds to the poles in the elastic $\xicp\db$ and $\xic\dvvb$ scattering processes.
The coupled-channel studies will be postponed until the next section.
The determination of the scattering length and effective range for each case is summarized in the columns fifth and sixth of Table~\ref{tab.ar}.

In the elastic $\xicp\db$ scattering, the $\pcs$ pole meets the working condition raised in Ref.~\cite{Guo:2015daa},
i.e., its mass is above the $\xicp\db$ threshold, so that the prescription in Eq.~\eqref{eq.x} is eligible for the probabilistic interpretation.
The small value of the compositeness $X$ of the $\xicp\db$ component, shown in the last column, indicates that 
other components  than the $\xicp\db$, play more important roles in the constitution of the $\pcs$.
The small value of $X$, \jo{whose central value is $31 \% $}, is also consistent with the rather large magnitude of $r$,
as the latter hints towards the appearance (within this simplistic elastic picture of $\Xi'_c\bar{D}$ scattering) 
of an underlying CDD pole near threshold, see Eq.~\eqref{eq.rcdd}.  
For the case of the elastic $\xic\dvvb$ scattering, the $\pcs$ pole 
does not meet the requirement of the working condition in Ref.~\cite{Guo:2015daa}, since the mass of the pole is below threshold. 
As a result, the prescription of Eq.~\eqref{eq.x} loses the probabilistic interpretation in the $\xic\dvvb$ case.
Nevertheless, the magnitudes of $r$ and $a$ all correspond to the conventional strong interactions,
which hint the natural explanation of the $\pcs$ as a molecular of the $\xic\dvvb$ system. 
This conclusion is further corroborated when discussing the coupled-channel case. 

 Note that the total width of the $\pcs$ is assumed to be saturated by only one channel, either the  $\xic\dvvb$ or $\xicp\db$, in the elastic ERE framework. This assumption should be revisited  since the partial decay width to the $\jpsi\Lambda$ channel is likely non-negligible, which is indeed the channel that is used to measure the $\pcs$ in experiment~\cite{Aaij:2020gdg}. As a result, a more realistic study would require the information of the {\it partial} decay width to $\xic\dvvb$ and $\xicp\db$, which will be investigated in the next section within a coupled-channel analysis. Later on various coupled-channel fits to the experimental event distributions of $\jpsi\Lambda$ will be also considered.

\section{Coupled-channel study of $\jpsi\Lambda$ and $\Xi_c\bar{D}^*$ by saturating the compositeness and the decay width }
\label{sec.threech}

There are intensive discussions in the literature to extend the Weinberg's compositeness relation  for the bound states \cite{Weinberg:1962hj,Weinberg:1965zz} to the resonance cases~\cite{Baru:2003qq,Hanhart:2011jz,Hyodo:2011qc,Aceti:2012dd,Sekihara:2014kya,Guo:2015daa,Gao:2018jhk,Matuschek:2020gqe}. The key obstacle in the latter situation is that the compositeness coefficient $X$ usually becomes complex \cite{Oller:2017alp}.
Different proposals have been given in Refs.~\cite{Hyodo:2011qc,Aceti:2012dd,Sekihara:2014kya,Guo:2015daa,Matuschek:2020gqe} to cure this problem.
In Ref.~\cite{Guo:2015daa} the Weinberg's compositeness relation for bound sates is generalized to resonances by performing a phase transformation of the $S$ matrix, so that the compositeness coefficient $X$ corresponding to the resonance state turns to be a positive number between 0 and 1, which can be interpreted as the probability to find the two-particle component in a resonance.
It is demonstrated in Ref.~\cite{Guo:2015daa} that the compositeness coefficient $X_j$ of a specific channel $j$ reduces to taking the absolute value of the product of the complex residue/coupling squared $g_j^2$ and the derivative of the two-point Green function at the resonance pole position $s_R=(M_R-i\Gamma_R/2)^2$~\cite{Guo:2015daa}, i.e.
\begin{eqnarray}\label{eq.xj}
X_j &=& |g_j|^2 \left| \frac{d G_j(s_R)}{d s} \right| \,, 
\end{eqnarray}
where $G_j(s)$ is the one-loop two-point Green function of the $j$th channel. We will take the explicit formula of $G_j(s)$ from the dimensional regularization method
\begin{eqnarray}\label{eq.gfunc}
G_j(s)  &=& -\frac{1}{16\pi^2}\left[ a(\mu^2) + \log\frac{m_2^2}{\mu^2}-x_+\log\frac{x_+-1}{x_+}
-x_-\log\frac{x_--1}{x_-} \right]\,, \nonumber\\
 x_\pm &=&\frac{s+m_1^2-m_2^2}{2s}\pm \frac{q_j(s)}{\sqrt{s}}\,,
\end{eqnarray}
where the subtraction constant $a(\mu^2)$ is introduced to replace the divergent term. 
The expression given in Eq.~\eqref{eq.gfunc} represents $G_j(s)$ in the first or  physical RS, and its corresponding formula \jo{in} the unphysical RS reads \cite{npa620}
\begin{eqnarray}\label{eq:G_RSII}
 G_j(s)^{\rm II} = G_j(s) - i \frac{q_j(s)}{4\pi\sqrt{s}}\,,
\end{eqnarray}
where $q_j(s)$ denotes the relativistic CM three-momentum of the $j$th channel
\begin{eqnarray}
q_j(s) =\frac{\sqrt{[s-(m_{1}+m_{2})^2][s-(m_{1}-m_{2})^2]}}{2\sqrt{s}}\,,
\end{eqnarray}
with $m_{1}$ and $m_{2}$ the masses of the two particles in that channel. 
The imaginary part of $G_j(s)^{\rm II}$ is opposite with the one of $G_j(s)$ along the real $s$ axis above the threshold. 
The explicit $\mu$ dependences in the first two terms in Eq.~\eqref{eq.gfunc} cancel each other, which implies that the $G_j(s)$ function is actually  independent on the scale $\mu$.
For the practical purpose, we will set $\mu=770$~MeV throughout. 
The natural value of the subtraction constant $a(\mu)$ can be estimated by matching the functions $G_j(s)$ calculated in dimensional regularization and with a three-momentum cut-off $q_\text{max}$ at threshold, as explained in Refs.~\cite{Oller:2000fj,Guo:2018tjx}, which leads to 
\begin{eqnarray}\label{eq.aval}
  a=-\frac{2}{m_1+m_2}\left[m_1\log\left(1+\sqrt{1+\frac{m_1^2}{q_{\rm max}^2}}\right)
  + m_2\log\left(1+\sqrt{1+\frac{m_2^2}{q_{\rm max}^2}}\right) \right] \simeq -2.5\,,
\end{eqnarray}
by taking $q_{\rm max}=1.0$~GeV, $m_1=m_{\Xi_c}$ and $m_2=m_{\dvv}$.  We will take a universal value for the subtraction constants in the two channels (since the variation in $a$ from Eq.~\eqref{eq.aval} by taking the masses of the particles in the first channel is small) and fix it to the one given in Eq.~\eqref{eq.aval}. 
 It is worth noticing that the derivatives of the $G_j(s)$ and $G_j(s)^{\rm II}$ functions appearing in the partial compositeness coefficient in Eq.~\eqref{eq.xj} do not depend on the subtraction constant $a(\mu)$ nor on $\mu$. 

We introduce the subscript $j=1$ and 2 to distinguish between the channels $\jpsi\Lambda$ and $\Xi_c\dvv$,
respectively. The saturation of the compositeness relation can be written as 
\begin{equation}\label{eq.gx}
 X= X_1 + X_2 =|g_1|^2 \left| \frac{d G_1^{\rm II}(s_R)}{d s} \right|+|g_2|^2 \left| \frac{d G_2(s_R)}{d s} \right|\,,
\end{equation}
where $X$ in the left-hand side is the total compositeness coefficient contributed by the $\jpsi\Lambda$ and $\xic\dvvb$ channels. 
Regarding the saturation of the decay width of $\pcs$, we also assume that it is exclusively given by these two channels. For the lighter case with $\jpsi\Lambda$, since its threshold is clearly below the resonance mass, one can safely use the standard two-body decay width formula~\cite{Zyla:2020zbs}
\begin{eqnarray}\label{eq.gamma1}
\Gamma_1=  |g_1|^2   \frac{q_1(M_R^2)}{8\pi M_R^2}\,. 
\end{eqnarray}
While for the heavier channel $\xic\dvvb$, its threshold is slightly above the resonance mass and the standard decay width formula can not be applied.
One way to estimate its partial decay width is to introduce the spectral distribution of the resonance taking into account its finite-width \cite{npa620,Meissner:2015mza,Kang:2016ezb} 
\begin{eqnarray}\label{eq.gamma2}
\Gamma_2 =  |g_2|^2 \int_{m_{\rm th,2}}^{M_R+n\Gamma_R} dw \,  \frac{q_2(w^2)}{16\pi^2 \,w^2}  \frac{\Gamma_R}{(M_R-w)^2+\Gamma_R^2/4} \,. 
\end{eqnarray}
Due to the long tail of the Lorentzian spectral distribution in the integrand of the above equation,
the upper integration limit should be taken with care to guarantee the coverage of the resonance peak in Eq.~\eqref{eq.gamma2}.
For the partial decay width of the $\pcs$ into the $\xic\dvvb$ channel,
the distance of its threshold to the $\pcs$ mass is 
slightly larger than the somewhat narrow width of the $\pcs$.
Therefore, somewhat large values of $n$ in the upper integration limit of Eq.~\eqref{eq.gamma2}
turn out to be required.
For example, it is found that when taking $n=8$, as addressed in detail in the next section, the decay width formulas for the near-threshold
channels $ \xic\dvvb$ and $\xicp\db$, calculated using Eq.~\eqref{eq.gamma2}, 
together with the standard formula of Eq.~\eqref{eq.gamma1} for $\jpsi\Lambda$,
 lead to consistent results from the pole determinations of both the total width
and the couplings fixed from the residues of the pole position. 
Consequently, we will take this value of $n=8$ in the following analyses.
The width saturation is then given by 
\begin{eqnarray}\label{eq.gwidth} 
\Gamma=\Gamma_1+\Gamma_2 = |g_1|^2  \frac{q_1(M_R^2)}{8\pi M_R^2} +  |g_2|^2 \int_{m_{\rm th, 2}}^{M_R+n\Gamma_R} dw \,\frac{q_2(w^2)}{16\pi^2 \,w^2} \frac{\Gamma_R}{(M_R-w)^2+\Gamma_R^2/4} \,.
\end{eqnarray}
Let us stress that we have also explicitly tried to set $n$ at larger values, say up to 50, and the results are found to be quantitatively similar to $n=8$. 
Therefore we conclude that the values presented in Tables~\ref{tab.swx1} and \ref{tab.swx2} are robust under the mild changes of $n$.

The equations that represent the saturation of the compositeness coefficient $X$~\eqref{eq.gx} and the decay width $\Gamma$ ~\eqref{eq.gwidth} can be used to determine the couplings $g_1$ and $g_2$, which in turn allow us to further analyze the partial compositeness factors and the partial decay widths. This approach based on saturating given values of $X$ and $\Gamma$ is not suitable to distinguish between the different total angular momenta $J$. The point is that the coupling strengths $|g_j|^2$ would appear in the same way for any total angular momentum $J$.  Possible contributions from lighter channels, such as the $\eta_c\Lambda$ channel, which threshold is rather distant from the interested energy region around 4.46~GeV, is effectively reabsorbed in the $\jpsi\Lambda$ channel.

To solve Eqs.~\eqref{eq.gx} and \eqref{eq.gwidth}, one must first provide the  value of the total compositeness $X$, which is however typically unknown a priori. Here we carry out exploring studies by giving some tentative values \jo{to} $X$, and the results are summarized in Table~\ref{tab.swx1} for the situation in which the $\jpsi\Lambda$ and $\xic\dvvb$ coupled channels are included,
and in Table~\ref{tab.swx2} for the $\jpsi\Lambda$ and $\xicp\db$ case.  

Three different values of $X$, namely $X=0.1, 0.5$ and 1.0, are used to analyze the $\jpsi\Lambda$ and $\xic\dvvb$ system. It turns out that the coupling strengths of $|g_1|$ and $|g_2|$ tends to decrease and increase, respectively, when increasing the values of $X$, so do the partial decay widths of $\Gamma_1$ and $\Gamma_2$. This trend is typical in this type of studies as previously observed \cite{Guo:2019kdc,Guo:2020pvt}. 
In all the cases, the partial compositeness factor $X_2$ of the $\xic\dvvb$ is always much larger than the compositeness coefficient $X_1$ contributed by the $\jpsi\Lambda$ channel.
Even when taking $X=1.0$, the magnitudes of the two partial widths of the $\jpsi\Lambda$ and $\xic\dvvb$ channels are similar because
$|g_2|\gg|g_1|$. 
While, when taking smaller values for the total compositeness coefficient $X$, the widths of the $\pcs$ are found to be dominated by the $\jpsi\Lambda$ channel. 

For the $\jpsi\Lambda$ and $\xicp\db$ coupled-channel system the maximum value of $X$ that results by solving Eqs.~\eqref{eq.gx} and \eqref{eq.gwidth} turns out to be around 0.3. According to the results shown in Table~\ref{tab.swx2},  the partial compositeness factor $X_2$ is always much larger than the value of $X_1$, regardless of the total compositeness coefficient $X$ taken in the study. In contrast, the magnitudes of the two partial widths are rather sensitive to the values of $X$. Larger values of $X$ lead to bigger partial widths $\Gamma_2$ of the $\xicp\db$ channel, and smaller $\Gamma_1$ for $\jpsi\Lambda$.

It is clear that definite values of $X$ will be obviously helpful to give more concrete conclusions of the properties of $\pcs$.
In the following section, we perform the fits within dynamical unitarized models to the experimental event distributions of the $\jpsi\Lambda$ from the LHCb collaboration~\cite{Aaij:2020gdg}. In this way, we are able to obtain more precise values of the coupling strengths $g_{i}$, which in turn allow us to perform predictions for the partial decay widths and compositeness coefficients contributed from different channels.

\begin{table}[htbp]
\centering
\begin{scriptsize}
\begin{tabular}{ c c c c c c c c c c}
\hline\hline
 & $|g_1|$ & $|g_2|$ & $\Gamma_1$ & $\Gamma_2$  & $X_1$   & $X_2$    \\
          &(GeV)  & (GeV)     & (MeV)      &  (MeV)    &      &    
\\ \hline
$X=0.1$   & $3.5_{-0.8}^{+1.0}$ & $4.3_{-0.4}^{+0.2}$ & $16.4_{-6.7}^{+9.5}$  & $0.9_{-0.5}^{+0.5}$ & $0.02_{-0.01}^{+0.01}$ & $0.08_{-0.01}^{+0.01}$   \\
\\ \hline
$X=0.5$   & $3.1_{-0.7}^{+0.7}$ & $10.4_{-0.8}^{+0.6}$ & $12.3_{-4.9}^{+5.9}$  & $5.0_{-2.9}^{+4.7}$ & $0.01_{-0.01}^{+0.01}$ & $0.49_{-0.01}^{+0.01}$   \\
\\ \hline
$X=1.0$   & $2.3_{-0.4}^{+0.4}$ & $14.8_{-1.0}^{+1.0}$ & $7.1_{-2.3}^{+1.7}$  & $10.2_{-5.5}^{+9.5}$ & $0.0_{-0.0}^{+0.0}$ & $1.0_{-0.0}^{+0.0}$   \\ \\
\hline\hline
\end{tabular}
\end{scriptsize}
\caption{Results from the $\jpsi\Lambda$ and $\xic\dvvb$ coupled-channel studies by assuming the saturation of the compositeness and decay width of the $\pcs$.\label{tab.swx1}  } 
\end{table}

\begin{table}[htbp]
\centering
\begin{scriptsize}
\begin{tabular}{ c c c c c c c c c c}
\hline\hline
  & $|g_1|$ & $|g_2|$ & $\Gamma_1$ & $\Gamma_2$  & $X_1$   & $X_2$    \\
          &(GeV)  & (GeV)     & (MeV)      &  (MeV)    &      &    
\\ \hline
$X=0.1$   & $3.2_{-1.0}^{+1.2}$ & $3.8_{-0.4}^{+0.2}$ & $13.0_{-6.4}^{+12.2}$  & $4.3_{-1.4}^{+1.8}$ & $0.01_{-0.00}^{+0.02}$ & $0.09_{-0.02}^{+0.00}$   \\
\\ \hline \\
$X=0.3$   & $1.4_{-0.0}^{+2.0}$ & $7.0_{-0.8}^{+0.4}$ & $2.5_{-0.0}^{+12.2}$  & $14.8_{-6.1}^{+4.3}$ & $0.00_{-0.00}^{+0.02}$ & $0.30_{-0.02}^{+0.00}$   \\ \\
\hline\hline
\end{tabular}
\end{scriptsize}
\caption{Results from the $\jpsi\Lambda$ and $\xicp\db$ coupled-channel studies by assuming the saturation of the compositeness and decay width of the $\pcs$. Notice that the maximum value of $X$ admitting solutions of the two equations~\eqref{eq.gx} and \eqref{eq.gwidth} is 0.3.
  \label{tab.swx2}} 
\end{table}

\section{Fits to the experimental event distributions}
\label{sec.fits}

In the following, we perform fits to the $\jpsi\Lambda$ invariant-mass distributions from the LHCb~\cite{Aaij:2020gdg} data on the decay $\Xi^-_b\to \jpsi\Lambda K^-$, in order to obtain the couplings and the resonance pole position. In this way one can further constrain the coupling strengths which will in turn give more definite predictions for $X$.

We consider first the $\Xi_c \bar{D}^*$ channel, and later also include the $\Xi'_c\bar{D}$ one, as these are the two-body channels whose thresholds, lying around 4478 and 4446~MeV, respectively, are very close to the resonance region of the $P_{cs}(4459)$. 
Because of this fact we study their interaction in $S$ wave only. 
In addition, we also include the much lighter channel $\jpsi\Lambda$, whose event distribution will be fitted. 
In principle one could also include the $\eta_c\Lambda$ since it is related to $\jpsi\Lambda$  by the HQSS, in which limit the mass of the $\jpsi$ and $\eta_c$ are equal. As we are interested only in the energy region around the $P_{cs}(4459)$ resonance it is then legitimate to characterize both in terms of just one effective channel, namely the $\jpsi \Lambda$ (more detailed discussions on this point are given below).

It is illustrative to consider the decomposition of the two-body states in the basis of HQSS eigenstates $|s_H\otimes j_\ell \ra$, where $s_H$ and $j_\ell$ are the total spin and angular momentum of the heavy and light quarks, respectively. The $|\Xi_c\ra$ and $|\bar{D}^{*}\ra$ spin multiplets are $|\frac{1}{2}\otimes 0 \ra_{1/2}$ and $|\frac{1}{2}\otimes \frac{1}{2}\ra_{1}$ in this basis, where the subscript indicates the spin $j_i$ of each state. The latter arises from the angular momentum combination of $s_{H_i}\otimes j_{\ell_i}$, with the index $i=1$ referring to $\Xi_c$ and $i=2$ to $\bar{D}^*$. One can then combine $j_1\otimes j_2$ in states of definite total angular momentum $J$, $|s_{H_1}j_{\ell_1}j_1;s_{H_2}j_{\ell_2}j_2;J\ra$, which
we simply designate in the following by $|\Xi_c\bar{D}^*\ra_J$.
Another way is to combine first  the heavy spins $s_{H_1}\otimes s_{H_2}$ and the light angular momenta $j_{\ell_1}\otimes j_{\ell_2}$ separately into states with well defined total heavy spin $s_H$ and total light angular momentum $j_L$, respectively.
Then, they are composed to give  states of the form $|s_{H_1} s_{H_2} s_H;j_{\ell_1} j_{\ell_2} j_L;J\ra$,
that we simply designate in the following by $|s_H\otimes j_L\ra_J$, with definite total angular momentum $J$.
 The relation between the two sets of basis vectors with  total angular momentum well defined is through the Wigner $9j$ symbols. See e.g. Appendix A of Ref.~\cite{Sakai:2019qph} for further details in this formalism.  In this way, it follows that
\begin{align}
\label{210708.1}
|\Xi_c\bar{D}^*\ra_{\frac{1}{2}}&= \frac{\sqrt{3}}{2} |0 \otimes \frac{1}{2} \ra_\frac{1}{2} 
- \frac{1}{2} |1 \otimes \frac{1}{2} \ra_\frac{1}{2}~,\\
|\Xi_c \bar{D}^* \ra_\frac{3}{2}& = |1 \otimes \frac{1}{2} \ra_\frac{3}{2}~.\nn
\end{align}
The states $|\jpsi\ra$, $|\eta_c\ra$ and $|\Lambda\ra$ are $|1\otimes 0\ra$, $|0\otimes 0\ra$ and
$|0\otimes \frac{1}{2}\ra$, in this order. Therefore, $|\eta_c\Lambda\ra=|0\otimes \frac{1}{2}\ra_\frac{1}{2}$, while $|\jpsi\Lambda\ra_J=|1\otimes \frac{1}{2}\ra_J$ has both $J=1/2$ and $3/2$. However, the transformation matrix for the $|\Xi_c\bar{D}^*\ra_{3/2}$ states is just the identity, as it is clear from Eq.~\eqref{210708.1}.

Here we only consider $S$-wave scattering because: i) The $|\Xi_c\bar{D}^*\ra$ and $|\Xi_c'\bar{D}\ra$ channels are very close to threshold in the region of interest. ii) The $|\jpsi\Lambda\ra$ ($|\eta_c \Lambda\ra$) does not couple to $|1\otimes \frac{1}{2}\ra$($|0\otimes\frac{1}{2}\ra$) in $D$ and higher partial waves  in the heavy-quark limit.

The coupled-channel scattering amplitudes take the form~\cite{Oller:1998zr,ollerbookrev}
\begin{eqnarray}\label{eq.ut}
 \mathcal{T}_J(s) = [\mathbb{I}-\mathcal{V}_J\cdot G(s)]^{-1} \cdot \mathcal{V}_J(s)\,.
\end{eqnarray}
The Eq.~\eqref{eq.ut} satisfies unitarity \cite{Oller:1998zr,Oller:2000fj} as the right-hand cut is generated by the diagonal matrix $G(s)$, whose elements are  $G_{j}(s)\equiv G(s,m_{j1}^2,m_{j2}^2)$, as defined in Eq.~\eqref{eq.gfunc}.

In all cases analyzed below, the matrix elements of $\mathcal{V}_J$ involving the $\jpsi$ or $\eta_c$ are taken constant because the energy region studied is narrow as compared with the energy distance to the thresholds of these channels. For the transitions  involving only the $|\Xi_c\bar{D}^*\ra$ or $|\Xi_c'\bar D\ra$ this will be also typically the case, except when we consider for comparison a possible linear dependence in $s$. 
Taking into account the conservation of the heavy-quark spin $s_H$ in the heavy-quark limit, and that $j_\ell=1/2$ always, we can write $V_J$ for $J=1/2$ and $3/2$ when explicitly including the $J/\psi\Lambda, \eta_c\Lambda$ and $\Xi_c\bar{D}^*$ channels as  
\begin{align}
\label{210711.1}
  \begin{array}{ll}
V_{\frac{1}{2}}=  \left(  \begin{array}{lll}
    0 & 0 & -\frac{1}{2}g\\
0 & 0 & \frac{\sqrt{3}}{2}g\\
-\frac{1}{2}g & \frac{\sqrt{3}}{2}g & C_\frac{1}{2}
\end{array}\right)\,,~
&
V_\frac{3}{2}=\left(
\begin{array}{lll}
  0 & 0 & g \\
  0 & 0 & 0 \\
  g & 0 & C_\frac{3}{2}
\end{array}
\right)~.
\end{array}
\end{align}
Though HQSS requires that $C_\frac{1}{2}=C_\frac{3}{2}$ we treat them separately because if significantly different values emerge when fitting data this could probably indicate that the model is incomplete. We have set to zero the direct transitions between the lighter channels $|\jpsi \Lambda\ra$ and $|\eta_c\Lambda\ra$ because they are Okubo-Zweig-Izuka suppressed. Furthermore, a recent lattice QCD study of the related $|\jpsi N\ra$ and $|\eta_cN\ra$ scattering  \cite{Skerbis:2018lew} found that these interactions are very weak. We also mention that the  coupled-channel model estimate of Ref.~\cite{Du:2020bqj} gives a $J/\psi p$ scattering length of the order of $10^{-3}$~fm. 

The typical formalism \cite{ollerbookrev} for implementing final-state interactions (FSI) and obtaining the needed amplitudes for the calculation of the actual event distribution of $\jpsi \Lambda$, that we call $F_J$, can be constructed by making use of the  matrix $D_J(s)$, 
\begin{align}
\label{210711.2}
D_J(s)&=\mathbb{I}-\mathcal{V}_J\cdot G(s)~.
\end{align}
 The matrix $D_J(s)$ appears in Eq.~\eqref{eq.ut} because its inverse resums the re-scattering effects, there starting from a basic interaction $V_J(s)$. But now the rescattering effects stem from the elementary production vertices $P_J$'s,
\begin{align}
\label{210711.4}
  P_J=\left( \begin{array}{l}
    0 \\
    0 \\
    d_J
  \end{array}
  \right)\,,
\end{align}
such that
\begin{align}
\label{210711.3}
F_J(s)&=D_J(s)^{-1} P_J\,.
\end{align}
To be consistent with the assumptions of Eq.~\eqref{210711.1}, the direct transition vertices with the $J/\psi\Lambda$ and $\eta_c\Lambda$ are also set to zero in Eq.~\eqref{210711.4}. 

However, the direct application of Eq.~\eqref{210711.3} for calculating $F_J(s)$ with a constant $P_J$ is not adequate because the production vertex
$F_J(s)$ is proportional to $G(s,M_{\Xi_c}^2,m_{D^*}^2)$, as it follows
from Eqs.~\eqref{210711.2}--\eqref{210711.3}.
But, this loop function  has a zero at an energy $E_z\simeq 4444~$MeV (below the $\Xi_c\bar{D}^*$ channel), just at the region of interest, 
so that $F_J(s)$ would also vanish at the same energy (something not seen in the data) unless  $P_J(s)$ has a pole there to cancel the zero.
 
 The simplest way to remedy this situation is to realize that the resonance signal  is actually contained in the determinant of $D_J(s)$, that we call
 $\Delta_J(s)$, since a resonance stems from a zero in it.
 The function $\Delta_J(s)$ also includes, of course, the cusp effects associated to the relevant nearby $\Xi_c\bar{D}^*$ threshold (and to the $\Xi_c' \bar{D}$ one when later included).\footnote{Another way would be to replace $d_J$ in $P_J$, Eq.~\eqref{210711.4}, by $d_J/(s-E_z^2)$. Nonetheless, we prefer to keep a constant $P_J(s)$, as in the original Eq.~\eqref{210711.4}.}
 As a result,  instead of Eq.~\eqref{210711.3}, the final expression that we use for the calculation of $F_J(s)$  is 
 \begin{align}
\label{210711.5}
F_J(s)&=\frac{d_J}{\Delta_J(s)}~.
 \end{align}

 Taking as in the HQSS limit that $G(s,M_{\jpsi}^2,M_\Lambda^2)=G(s,M_{\eta_c}^2,M_\Lambda^2)$, which is actually a good approximation because the region under study is much higher than the $\jpsi\Lambda$ and $\eta_c\Lambda$ thresholds, we have that both $\Delta_{1/2}$ and $\Delta_{3/2}$ become equal to
 \begin{align}
\label{210803.1}
\Delta_{J}=1-G_{3}(C_{J}+G_{1} g^2)~.
 \end{align}
 Because of these reasons in the following we do not take explicitly into consideration the $\eta_c \Lambda$ channel and only keep the $\jpsi \Lambda$ one. As a result, the number of coupled channels is reduced to two. 
 
The invariant-mass distribution of the $\jpsi \Lambda$, $dN(s)/d\sqrt{s}$, is calculated from $F_J(s)$ by implementing the three-body phase space as 
\begin{align}
\label{210803.2}
\frac{dN(s)}{d\sqrt{s}}&=\frac{1}{128\pi^3 M^3_{\Xi_b}}\frac{\sqrt{\lambda(M^2_{\Xi_b},s,M_K^2)\lambda(s,M^2_{\jpsi},M_\Lambda^2)}}{\sqrt{s}}\sum_{J}|F_J|^2~.
\end{align}
In addition, we take into account the width of the experimental bin by convoluting the signal with a function to consider the effects of the spread in energy, typically with a Gaussian whose $\sigma$ is 2.6~MeV (around half of the bin size).

For every $T$ matrix obtained in our studies we calculate the resonance pole positions and their residues. 
The resonance poles lie in the complex energy plane of an unphysical RS, which can be accessed via the analytical extrapolation of the $G_j(s)$ functions. The expression given in Eq.~\eqref{eq.gfunc} represents $G_j(s)$ in the first or physical RS, and the corresponding formula on its unphysical RS 
is given in Eq.~\eqref{eq:G_RSII}  
with the square root calculated in its first RS (so that the phase of its argument is taken between 0 and $2\pi$ in the complex $s$ plane). 
The imaginary part of $G_j(s+i\epsilon)^{\rm II}$ is opposite to the one of $G_j(s+i\epsilon)$ for $s$ taking physical values (real ones above threshold), with $\epsilon\to 0^+$.
Different unphysical RSs of the coupled-channel scattering amplitudes in Eq.~\eqref{eq.ut} can be accessed by properly taking $G_j(s)$ or $G_j(s)^{\rm II}$ for different channels. The second RS can be labeled as $(-,+)$, where the plus(minus) sign in the $j$th entry  indicates taking $G_j(s)$($G_j(s)^{\rm II})$ in the $j$th channel. In this convention, the first, third, and  fourth Riemann sheets are labeled as $(+,+)$,  $(-,-)$, and $(+,-)$, respectively. The most relevant resonance poles are found to lie in the  second RS, which connects continuously with the physical RS below the $\Xi_c\bar D^\ast$ threshold.

The matrix elements of the  scattering matrix in the unphysical RS around the resonance pole position can be written as 
\begin{eqnarray}
  \label{eq.res}
\left.\mathcal{T}_J\right|_{kj}(s) = -\frac{g_k g_j}{s-s_{\rm pole}} + \cdots \,,
\end{eqnarray}
where $g_{k,j=1,2}$ are the effective couplings of the resonance to the corresponding channels. Here, we follow the usual procedure, perfectly justifiable for narrow resonances,  of
identifying $s_{\rm pole}=(M_R-i\Gamma_R/2)^2$. 
  The omitted terms in Eq.~\eqref{eq.res} are the regular parts in the Laurent expansion around  $s_{\rm pole}$.


\subsection{Fits with the $\jpsi\Lambda$ and $\Xi_c\bar{D}^*$ channels}
\label{sec.210803.2}

First we present fits for $dN(s)/d\sqrt{s}$ including only the  $\jpsi\Lambda$ and $\Xi_c\bar{D}^*$ channels
in which we add progressively $J=1/2$ and then $3/2$ (both required by HQSS, cf. Eq.~\eqref{210708.1}).
We also separate the discussion between the cases in which only a constant potential is used and when we also
consider a linear $s$ dependence.

Firstly, the potential is taken as the constant matrix 
\begin{align}
\label{210902.1}
  V_\frac{1}{2}&=\left(
  \begin{array}{ll}
    0 & g \\
    g & C_\frac{1}{2}
  \end{array}
  \right)~,
\end{align}
and we use Eq.~\eqref{210711.5} with $d_\frac{1}{2}$ for $J=1/2$. 
Here we follow two strategies. 
One of them consists of fixing the only two free parameters \jo{in $V_\frac{1}{2}$} by reproducing the experimental $P_{cs}(4459)$ resonance pole in $1/\Delta_{1/2}$ \jo{in} the second RS.
The other strategy is to fit directly the data and obtain $g$, $C_{1/2}$ and $d_{1/2}$.

The equation that results by imposing $\Delta_{1/2}(s_R)=0$ is directly obtained from Eq.~\eqref{210803.1} 
by taking $G_{1}(s)$ evaluated in its second RS, $G_1^{\rm II}(s)$, with $s_R=(M_R-i\Gamma_R/2)^2$ determined from the experimental central values,
 $M_R=4458.8$ MeV and $\Gamma_R =17.3$~MeV \cite{Aaij:2020gdg}. 
The resulting event distribution is shown by the dot-dashed line in the top left panel of 
Fig.~\ref{fig.210803.1}, and we see that it provides a  qualitative  agreement with the experimental data.

Now, we proceed with the direct fits to data by minimizing a standard $\chi^2$. 
We show the fit results in Fig.~\ref{fig.210803.1} and provide the fitted values of the parameters in Table~\ref{tab.210803.1}.
Every type of fit is described in the left-most column of the table by $(abc)$ where: $a$ refers to the number of channels, 2 ($J/\psi\Lambda$-$\Xi_c\bar{D}^*$ or $J\psi\Lambda$-$\Xi^\prime_c\bar{D}$) or 3 ($J/\psi\Lambda$-$\Xi^\prime_c\bar{D}$-$\Xi_c\bar{D}^*$ described in Sec.~\ref{sec.210804.1}); $b$ to the number of partial waves, 1 ($J=1/2$ or $3/2$) or 2 ($J=1/2$ and $3/2$); and $c$ indicates whether we use a constant potential (0) or include a linear dependence in $s$ (1) for the near-threshold coupled-channel potential.
We use a cross when an entry is not meaningful to be given  (typically in those instances with only 2 channels involved).
For each fit we also give the position of the resonance pole  and the residues of the $T$ matrix, collected in Table~\ref{tab.210803.2}, following the same conventions as just explained. The energy region actually fitted is shown by the shaded area in Fig.~\ref{fig.210803.1}, comprising eleven experimental points around the $P_{cs}(4459)$ peak.

In the first panel of Fig.~\ref{fig.210803.1} the fit   $(210)$ corresponds to the solid line, which provides a good overall fit to data given the experimental uncertainties. Nonetheless, we also notice that it lies somewhat low compared to data in the region to the left of the peak. The values of the fit parameters are given in Table~\ref{tab.210803.1}. 

We have also tested the case of allowing a linear $s$ dependence in the $\Xi_c\bar{D}^*$  interacting channels, so
that $C_\frac{1}{2}\to C_\frac{1}{2}(\frac{s}{M_{\rm CDD}^2}-1)$ in Eq.~\eqref{210902.1}.
The position of the zero is called $M_{\rm CDD}^2$, because
in the case of the $\Xi_c\bar{D}^*$ one-channel scattering it would be a CDD pole \cite{castillejo,Oller:1998zr}. However,
 the solution found with the smallest $\chi^2$ always gives rise to a pole in the first or physical RS, which is not admissible. This is related to the fact that these fits of smallest $\chi^2$ produce $C_\frac{1}{2}<0$, which allows the vanishing of the denominator of the one-channel $T$ matrix,\begin{align}
\label{210803.3}
\frac{1}{C_{1/2}(s/M_{\rm CDD}^2-1)}+G_{2}(s)~.
\end{align}
 This can be easily seen because  $\Im G_2(s)<0$ for $\Im s>0$ in the first RS, and then it has opposite sign to $s^*/C_\frac{1}{2}$ for $C_\frac{1}{2}<0$. The absence of an acceptable solution with a nearby CDD pole is in favor of interpreting the $P_{cs}(4459)$ as a dynamically generated resonance, i.e., a $\xic\dvvb$ molecular state.

Next, we include together the two angular momenta $J=1/2$ and $3/2$,
and then we have two more free parameters $C_\frac{3}{2}$ and $d_\frac{3}{2}$,
\begin{align}
\label{210804.2}
V_\frac{3}{2}=\left(
\begin{array}{ll}
  0 & g\\
g & C_\frac{3}{2}
\end{array}
\right)~,~
P_\frac{3}{2}&=\left(\begin{array}{l}
  0 \\
  d_\frac{3}{2}
\end{array}
\right)~. 
\end{align}
According to Eq.~\eqref{210803.2}
we have now $|F_\frac{1}{2}(s)|^2+|F_\frac{3}{2}(s)|^2$. Notice, that both $F_\frac{1}{2}$ and $F_\frac{3}{2}$ obey the same type of
expression, so that they are interchangeable. This is important when listing the poles found for the $(220)$ fit because it is then not really possible to discern which is actually the angular momentum of each of them.
The results of the fit are given by the entry $(220)$ in Tables~\ref{tab.210803.1} and \ref{tab.210803.2}. The
resulting $\chi^2=2.95$ is much lower than for the fit of type $(210)$, and this is clearly observed by the
 curve obtained for $(220)$ in the second panel of Fig.~\ref{fig.210803.1}, which runs much closer to the central values 
 of data and there are now two peaks corresponding to two poles, listed in Table~\ref{tab.210803.2}. However, there is the caveat that $C_\frac{1}{2}$ and $C_\frac{3}{2}$ are very different and are not compatible. The large deviation between the $C_\frac{1}{2}$ and $C_\frac{3}{2}$ can not be simply 
 accounted for by the HQSS breaking which implies their difference should be of order of $\mathcal{O}(\frac{\Lambda_\text{QCD}}{m_c})$. 
 This issue is a strong reason for
 exploring three-channel scattering including the $\jpsi\Lambda$, $\Xi_c\bar{D}^*$ and $\Xi_c'\bar{D}$ states,
 as we do in the Sec.~\ref{sec.210804.1}.


\subsection{Fits including only the $\jpsi\Lambda$ and $\Xi'_c\bar{D}$ channels}
\label{sec.210803.1}

The decomposition of the $|\Xi_c'\bar{D}\ra$ in the HQSS basis is
\begin{align}
\label{210803.4}
|\Xi_c'\bar{D}\ra&=-\frac{1}{2}|0\otimes\frac{1}{2}\ra_\frac{1}{2}'
+\frac{\sqrt{3}}{2}|1\otimes\frac{1}{2}\ra_\frac{1}{2}'~,
\end{align}
and we can only have the $J=1/2$ contribution in this scenario. When referring to this case we include primes to indicate that $\Xi_c'\bar{D}$ is considered instead of $\Xi_c\bar{D}^*$. 

We have the same number of free parameters as when including the $\jpsi\Lambda$ and $\Xi_c\bar{D}^*$ states.
However, when the $\jpsi\Lambda$ and $\Xi_c'\bar{D}$ are the ones considered instead the resulting fits
with either constant potentials or allowing a linear $s$ dependence in the $\Xi'_c\bar{D}$ direct transition potential are not able to reproduce the data even qualitatively.
This is clear from the resulting curve shown in the third panel of Fig.~\ref{fig.210803.1}, where the 
curve labeled by ``Cons" and ``CDD" correspond to the constant potentials and allowing a linear $s$ 
dependence, respectively. 
Therefore, in light of the results here and in Sec.~\ref{sec.210803.2} one can clearly conclude that the $P_{cs}(4459)$ is dominated by the $\Xi_c\bar{D}^*$ interactions, and much less by the  $\Xi_c'\bar{D}$ channel.  This is also in agreement with the results in Secs.~\ref{sec.ere} and \ref{sec.threech}.


\subsection{Fits including the three channels $\jpsi\Lambda$, $\Xi_c'\bar{D}$ and $\Xi_c\bar{D}^*$} 
\label{sec.210804.1}

All the  three channels  $\jpsi\Lambda$ (1), $\Xi_c'\bar{D}$(2), and $\Xi_c\bar{D}^*$(3), 
with the channel numbers given between brackets, 
interact among them only in $J=1/2$.
The interaction potential $V_\frac{1}{2}$ is taken as
\begin{align}
\label{210804.1}
V_\frac{1}{2}&=\left(
\begin{array}{lll}
0 &  g'     & g        \\ 
 g'  & C'_\frac{1}{2} &  C_{\rm mx} \\
g & C_{\rm mx}     &  C_\frac{1}{2}
\end{array}
  \right)~,
\end{align}
where all the matrix entries are constants. We also have the elementary production vertices
$P_\frac{1}{2}=(0\,,  d_\frac12^\prime \,, d_\frac{1}{2})$.
 The $J=3/2$ potential is left formally unchanged, cf. Eq.~\eqref{210804.2}, but now having included the three channels
we impose the HQSS constraint $C_\frac{1}{2}=C_\frac{3}{2}$. Once the $\Xi^\prime\bar{D}$ is explicitly included, it could provide a mechanism for the mass split of the two $P_{cs}$ resonant states. That is, we assume that the main source of HQSS breaking in the
relation $C_\frac{1}{2} \ne C_\frac{3}{2}$ in Sec.~\ref{sec.210803.2}, fit $(220)$, is due to the neglected nearby $\Xi'_c\bar{D}$ channel. All in all, we have 8 free parameters, $g$, $g'$, $C_\frac{1}{2}$, $C_\frac{1}{2}'$, $C_{\rm mx}$, $d_\frac{1}{2}$, $d_\frac12^\prime$ and
$d_\frac{3}{2}$. 

\begin{figure}[htbp]
\centering
\begin{tabular}{ll}
\includegraphics[width=0.45\textwidth,angle=-0]{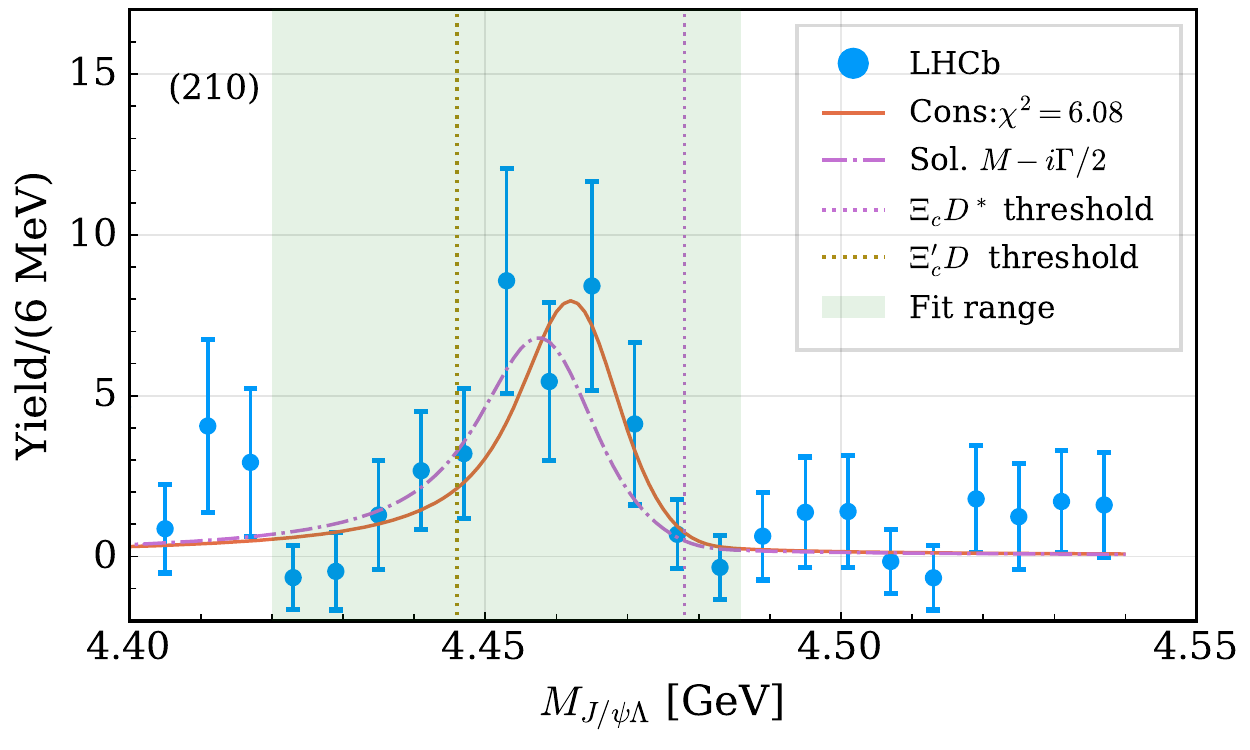} 
& \includegraphics[width=0.45\textwidth,angle=-0]{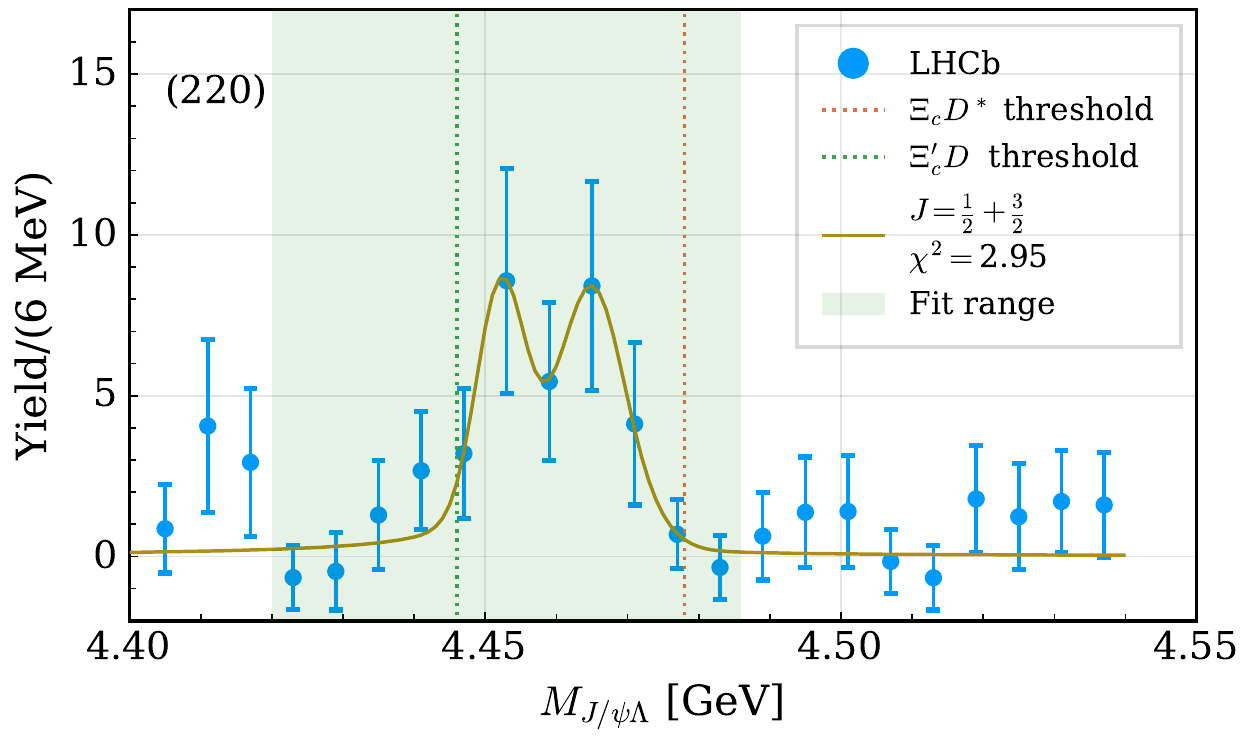}\\
\includegraphics[width=0.45\textwidth,angle=-0]{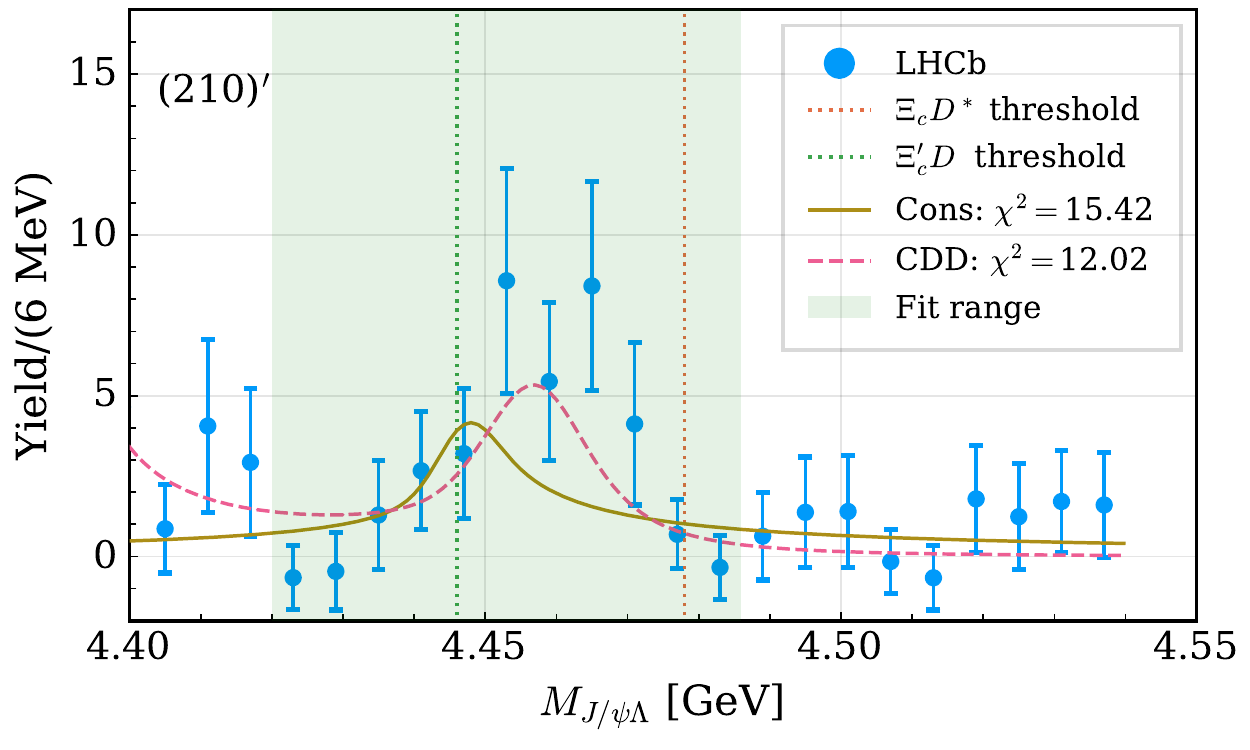} 
& \includegraphics[width=0.45\textwidth,angle=-0]{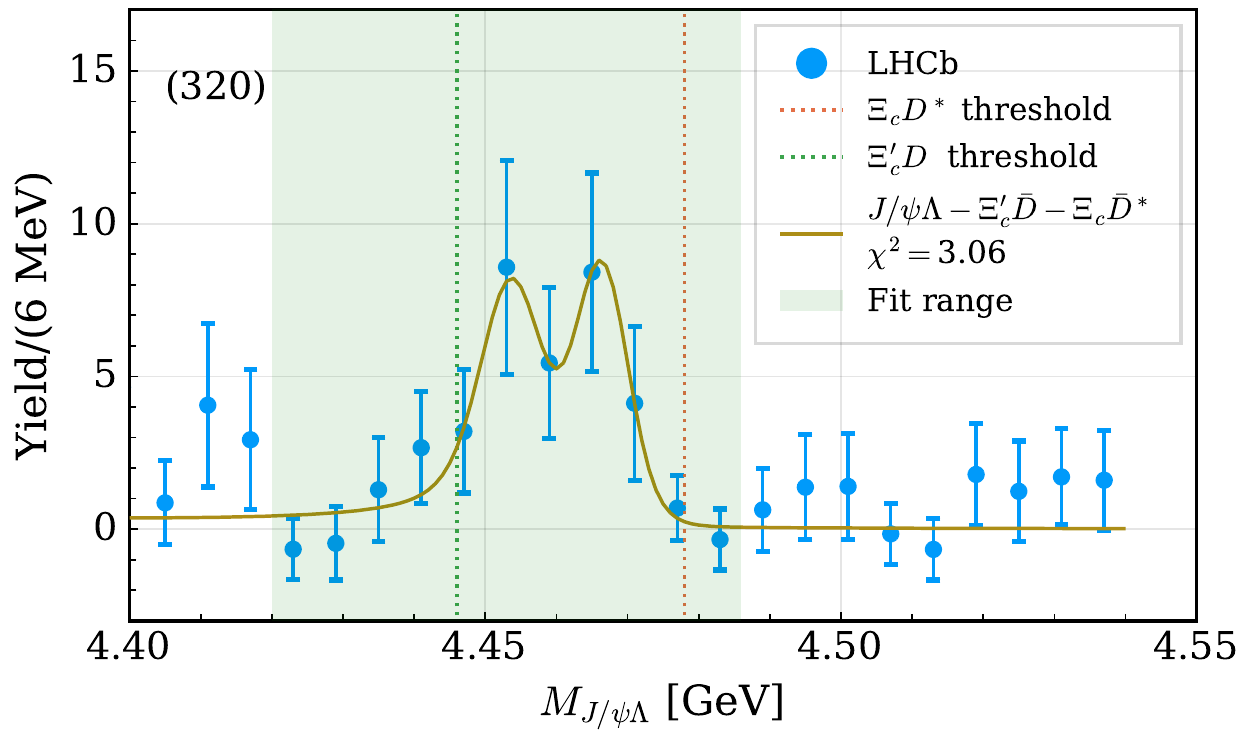} 
\\
\end{tabular}
\caption{The panels are ordered from top to bottom and left to right. The solid lines (and dashed line in the third figure) give the result of the fits, and the type of 
fit is indicated at the right top corner of every panel.  The greenish area singled out in the figures corresponds to the energy region actually fitted.
  We also indicate by the vertical lines the thresholds from left to right of the $\Xi_c'\bar{D}$ and $\Xi_c\bar{D}^*$, respectively. 
  The experimental data from Ref.~\cite{Aaij:2020gdg} are the points with errorbars.
 \label{fig.210803.1}} 
\end{figure}

We explore first a perturbative inclusion of the $\Xi'_c\bar{D}$ channel so that
we take $g'=C'_\frac{1}{2}=d'_\frac{1}{2}=0$, and then its inclusion is similar to that for the $\jpsi\Lambda$ channel, though
$G_2(s)$ is peaked in the $\Xi'_c\bar{D}$ threshold. We then obtain the fit $(320)$.
 Even though we now impose that $C_\frac{1}{2}=C_\frac{3}{2}$, the $\chi^2=3.06$ is almost equal to
that for the fit $(220)$, and the fit (320) reproduces very well the data as shown in the last panel of Fig.~\ref{fig.210803.1}. The values for the $\chi^2$ and the fit parameters are given in Tables~\ref{tab.210803.1} and \ref{tab.210804.1a}. 
In addition we also notice that $C_\frac{1}{2}=C_\frac{3}{2}$ is almost identical to the values for $C_\frac{1}{2}$ registered for the fits
$(210)$ and $(220)$, which then indicates that this parameter is well determined within our approach.
Related to this, we conclude then that our hypothesis that the lack of the channel $\Xi'_c\bar{D}$ is the reason for the
  HQSS breakdown in the relation $C_\frac{1}{2}=C_\frac{3}{2}$, as obtained in the fit (220),  is well confronted to phenomenology. 

The nomenclature for the RSs with three channels is arranged in increasing values for the thresholds.  
Namely, the RSII $(-++)$ is the one connecting with the RSI $(+++)$ above the $\jpsi\Lambda$ and below the $\Xi_c'\bar{D}$ thresholds,
the RSIII $(--+)$ connects with the physical sheet  between the $\Xi_c'\bar{D}$ and  the $\Xi_c\bar{D}^*$ thresholds, and
the RSIV connects with RSI above the $\Xi_c\bar{D}^*$ threshold.  In this case the scattering in the $J=1/2$ and $J=3/2$ angular
momenta is not interchangeable and we can give the corresponding spin of the resonances found in Table~\ref{tab.210804.1}.

The two poles observed for the fit $(320)$, collected in Table~\ref{tab.210804.1}, one with $J=1/2$ (first peak in the last panel of Fig.~\ref{fig.210803.1}) and another with $J=3/2$ (the higher one in the same plot), 
have each of them an imaginary part that is  smaller in absolute value than the one in the first two poles of Table~\ref{tab.210803.2}.  
This is indeed related with the size of the coupling $g$ which is also  smaller for the $(320)$ fits as compared with the $(220)$.
Since this coupling controls how strongly the 
relatively-light channel $\jpsi\Lambda$ couples to $\Xi_c\bar{D}^*$, and the resonances are strongly affected by the latter channel, the smaller the value of
$g$ the smaller the resulting widths because the residue $g_1$ is proportional to it.

We have also explored new fits in which $g'$, $C'_\frac{1}{2}$ and/or $d'_\frac{1}{2}$ are allowed to float. However, we do not observe any significant change, either qualitative or quantitative, in the description of the data. For instance, the resulting $\chi^2$ in every case barely improves compared to the fit $(320)$. 
Of course, the resulting fits are typically over-determined so that some of the parameters are completely undetermined (having errorbars that are much bigger than their central values).

Let us discuss two of such fits.
If we still fix $C'_\frac{1}{2}=0$ but release $g'$ we have the fit $(320)_1$ in Tables~\ref{tab.210803.1} and \ref{tab.210804.1a}.  We observe only a slight improvement in the $\chi^2$, $g'=59.6^{+160.7}_{-249.2}$ is compatible with zero and basically undetermined, while the rest of fit parameters are very similar to the reference fit with three channels $(320)$. However, if we fix $g'=0$ and release $C'_\frac{1}{2}$ the resulting fit $(320)_2$ in Tables~\ref{tab.210803.1} and \ref{tab.210804.1a}, gives rise to a very  uncertain coupling $g$.
Though, it is still compatible within errors with the determinations from the other three-channel fits,
its central value is very different, while for the other  fits $(320)$ and $(320)_1$ their central values are quite similar. 
Notice also that again the $\chi^2=2.81$ has improved only by a basically negligible amount compared to $(320)$.
The pole content for these fits is not shown because it is very similar to the one already given in Table~\ref{tab.210804.1}. 

In summary, our reference fit for the three-channel case is the $(320)$ one, in which the $\Xi'_c\bar{D}$ channel is introduced perturbatively ($g'=C'_\frac{1}{2}=d'_\frac{1}{2}=0$) and we also take into account both $J=1/2$ and $J=3/2$. Releasing  either $g'$ or $C'_\frac{1}{2}$ does not really provide better fits and the values for the fit parameters are compatible with those from $(320)$. This indicates a perturbative role of the $\Xi_c'\bar{D}$ channel, which is also in agreement with our previous results  in Sec.~\ref{sec.210803.1} when considering only the $\jpsi\Lambda$ and $\Xi'_c\bar{D}$ channels in $J=1/2$. Let us recall that \jo{ in that case} it was not possible to account for the $P_{cs}(4459)$ resonance signal.

\begin{table}
  \begin{center}
    \begin{tabular}{lllllll}
Fit   & $\chi^2$ & $g$                 & $C_\frac{1}{2}$            & $d_\frac{1}{2}$            &  $C_\frac{3}{2}$ & $d_\frac{3}{2}$ \\
\hline
&&&&&&\\
$(210)$ & 6.08  & $316.0^{+86.2}_{-69.3}$ & $1126.6^{+327.8}_{-214.9}$ & $290.5^{+71.1}_{-77.2}$ & $\times$ & $\times$  \\
&&&&&&\\
$(220)$ & 2.95  & $217.8^{+80.6}_{-79.3}$ & $1125.5^{+190.6}_{-185.9}$ & $174.7^{+86.5}_{-77.3}$   & $3862.7^{+1466.1}_{-1003.3}$ & $97.6^{+37.9}_{-35.5}$ \\
&&&&&&\\
$(320)$ & 3.06  & $124.5^{+130.4}_{-164.7}$ & $1105.8^{+191.9}_{-132.5}$ & $250.8^{+62.0}_{-40.2}$ & $C_\frac{3}{2}=C_\frac{1}{2}$ & $82.1^{+292.1}_ {-128.0}$\\ 
&&&&&&\\
$(320)_1$ & $3.00$ & $145.8^{+122.9}_{-414.5}$ & $1108.7_{-142.9}^{+192.3}$ & $237.8^{+84.6}_{-72.5}$ &
$C_\frac{3}{2}=C_\frac{1}{2}$ & $100.1_{-144.2}^{+125.1}$ \\
&&&&&&\\
$(320)_2$ & $2.81$ & $0.06^{+289.5}_{-288.6}$ & $1098.0^{+175.7}_{-137.5}$ & $197.9_{-78.3}^{+177.8}$ &
$C_\frac{3}{2}=C_\frac{1}{2}$ & $19.1^{+90.2}_{-90.2}$ \\
&&&&&&\\
\hline
    \end{tabular}\caption{Coupling parameters in $V_\frac{1}{2}$ and $V_\frac{3}{2}$ related to $\Xi_c\bar{D}^*$ for several fits.
      The type of fit and the $\chi^2$ are also given in the first and second columns, respectively. For the meaning of different fits, see the text for details. Note that the values of $C_{\frac12}$($d_{\frac12}$) and $C_\frac32$($d_\frac32$) are interchangeable in Fit (220). \label{tab.210803.1}}
\end{center}
  \end{table}

\begin{table}
  \begin{center}
  \begin{tabular}{lllll}
    Fit & RS    & $\sqrt{s_R}$   (MeV)&  $|g_1|$   (MeV) & $|g_2|$   (MeV)\\
    \hline
    &&&\\
    $(210)$ & $(-+)$ &   $4463.2^{+ 2.8}_{-4.4}-i\,7.1^{+2.5}_{-2.8}$ & $3.29^{+0.64}_{-0.68}$  & $13.81^{+0.87}_{-0.68} $ \\
    &&&\\
    $(220)$ & $(-+)$ &  $4465.5^{+ 2.3}_{-2.3}-i\,3.8^{+2.3}_{-3.4}$  & $1.20^{+0.46}_{-0.44} $ & $13.01^{+0.65}_{-0.62}$ \\
    &&&\\
   $(220)$ & $(-+)$ &  $4452.1^{+ 2.4}_{-2.0}-i\,0.5^{+0.3}_{-0.7}$  & $0.88^{+0.46}_{-0.33}$ &  $15.73^{+0.33}_{-0.41}$ \\
      &&&\\
  \hline
  \end{tabular}\caption{Poles in the second RS for the studies with the $\jpsi\Lambda$ and $\Xi_c\bar{D}^*$ channels included. 
\label{tab.210803.2}}
\end{center}
  \end{table}

\begin{table}
  \begin{center}
    \begin{tabular}{llllll}
Fit  & $\chi^2$ &  $C_{\rm mx}$  &  $g'$                 & $C'_\frac{1}{2}$            & $d'_\frac{1}{2}$   \\
\hline
&&&&&\\
$(320)$ & 3.06  & $851.1^{+341.0}_{-148.1}$ & 0 & 0 & 0  \\
&&&&&\\
$(320)_1$ & 3.00 & $885.6^{+353.4}_{-204.9}$ & $59.6^{+160.7}_{-249.2}$ & 0 & 0\\
&&&&&\\
$(320)_2$ & 2.81 & $579.8^{+1014.2}_{-2201.5}$ & $0$ & $187.6^{+174.8}_{-831.8}$ & 0\\
&&&&&\\
\hline
    \end{tabular}\caption{Coupling parameters in $V_\frac{1}{2}$ related to the $\Xi_c'\bar{D}$ channel in the three-coupled channel fits.
      \label{tab.210804.1a}}
\end{center} 
  \end{table}

\begin{table}
  \begin{center}
  \begin{tabular}{lllllll}
    Type & $J$ & RS    &  $\sqrt{s_R}$ & $|g_1|$ & $|g_2|$ & $|g_3|$ \\
         &     &       &   (MeV)       &  (MeV)     &  (MeV)  & (MeV)  \\
    \hline
    &&&&\\
$(320)$ & 3/2 & RSII   &  $4466.6^{+ 1.9}_{-2.7}-i\,1.3^{+1.3}_{-3.7}$ &  $1.4^{+1.4}_{-1.4}$  & $\times$ & $12.6^{+0.8}_{-0.6}$\\
    &$(-+)$&&&\\
    &&&&\\
$(320)$ & 1/2 & RSIII  &  $4453.8^{+ 2.4}_{-3.3}-i\,2.8^{+0.9}_{-0.8}$ &  $0.6^{+0.6}_{-0.6} $ &  $4.2^{+0.2}_{-0.4}$ & $15.0^{+0.5}_{-0.3}$  \\
    &$(--+)$&&&\\
    &&&&\\
    \hline
  \end{tabular}\caption{The poles for the fit $(320)$ with three channels. Only the poles lying in the energy region fitted and in the RSs connected to the
    physical sheet are presented. These are the poles that reflect in the peaks visible in the fits. We also gather the couplings to all channels included. For the meaning of different fits and RSs, see the text for details. \label{tab.210804.1}}
\end{center}
  \end{table}

\subsection{The three-coupled channel fit (320) and the revisit of saturation of the width and compositeness}
\label{sec.210902.1}
 
With the determined pole parameters in Table~\ref{tab.210804.1} for the fit (320) involving the three-coupled channels $\jpsi\Lambda$, $\xicp\db$ and $\xic\dvvb$, we can revise the discussions of the compositeness and partial decay widths, and make comparisons with the studies based on the saturation of the compositeness coefficient and decay width in the Sec.~\ref{sec.threech}. For the partial decay width of the $\pcs$ into the $\jpsi\Lambda$ channel, one can safely use the standard decay width formula in Eq.~\eqref{eq.gamma1}. For the partial widths into the near-threshold channels $\xicp\db$ and $\xic\dvvb$, we will rely on the Lorentzian spectral integration formula in Eq.~\eqref{eq.gamma2}. 
By taking $n=8$, the partial decay widths for the lighter resonance pole with $J=1/2$ in Table~\ref{tab.210804.1} are found to be 
\begin{eqnarray}
\Gamma_1= 0.5_{-0.5}^{+1.9}~{\rm MeV}\,, \quad \Gamma_2= 4.3 _{-1.4}^{+1.2}~{\rm MeV}\,, \quad \Gamma_3= 0.9_{-0.6}^{+1.2}~{\rm MeV}\,, 
\end{eqnarray}
and the sum of the three widths agrees nicely with the width of the corresponding resonance pole, i.e. minus twice the imaginary part of the pole. This is also the main reason to take $n=8$ here as well as in Sec.~\ref{sec.threech}.
The partial compositeness coefficients contributed by the $\jpsi\Lambda$ and $\xicp\db$ based on the pole parameters in Table~\ref{tab.210804.1} are 
\begin{eqnarray}
X_1= 0.0\pm 0.0\,, \quad  X_2= 0.15\pm 0.05 \,.
\end{eqnarray}
It should be noted that for the $\xic\dvvb$ channel the pole position of the lighter resonance state with $J=1/2$ does not meet the working condition of Ref.~\cite{Guo:2015daa}, and as a result it is not very meaningful to interpret the partial  $X_2$ (corresponding to $X_3$ in the three-channel case) calculated in Eq.~\eqref{eq.xj} as the probability of the $\xic\dvvb$ component in that resonance pole. In fact, had we blindly used Eq.~\eqref{eq.xj} to calculate $X_2$ its value found would be 0.94, so that the sum of compositeness for the three channels would turn out to be slightly larger than 1. This is a manifestation that one should follow the working conditions argued in Ref.~\cite{Guo:2015daa} for giving a probabilistic interpretation of the compositeness $X$ of a resonance.

For the heavier resonance pole with $J=3/2$ in Table~\ref{tab.210804.1} , the partial widths calculated for the $\jpsi\Lambda$ and $\xic\dvvb$ are  
\begin{eqnarray}
 \Gamma_1= 2.6_{-2.6}^{+8.2}~{\rm MeV}\,, \quad \Gamma_3= 0.4_{-0.4}^{+2.5}~{\rm MeV}\,,
\end{eqnarray}
and the resulting partial compositeness coefficients turn out to be 
\begin{eqnarray}
 X_1= 0.0\pm 0.0\,, \quad  X_3= 1.0_{-0.2}^{+0.2}\,.
\end{eqnarray}
Despite that the mass of the $J=3/2$ pole lies clearly below the threshold of $\xic\dvvb$ we have given $X_3$ calculated from Eq.~\eqref{eq.xj} as an indication that this resonance is clearly dominated by the $\xic\dvvb$ channel component. A similar comment could be also made in relation to $X_3$ for the $J=1/2$ pole. Indeed, we also find for both poles that the coupling $|g_3|$ is huge in comparison with $|g_1|$ and much larger, by around a factor 7/2, than $|g_2|$ for $J=1/2$.  
We point out that a direct comparison of the results for the partial widths and compositeness coefficients in this section and the ones in Sec.~\ref{sec.threech} is not straightforward. The point is that in Sec.~\ref{sec.threech} the discussions are based on  the masses and widths of the $\pcs$ from the single-state determination of the LHCb collaboration~\cite{Aaij:2020gdg}. However, in the present section our reference fit prefers the two-resonance solutions and the widths of the two resonance poles taking separately are found to be much smaller than the single-state determinations from the LHCb, though the combination of the two nearby peaks, convoluted with the experimental resolution in energy, gives rise to a total signal as a double bump with a width in agreement with the one pole determination for the $P_{cs}(4459)$ \cite{Aaij:2020gdg}. Nonetheless,  both the results here and in Sec.~\ref{sec.threech} are essentially compatible, as one concludes by comparing the couplings $|g_i|$ in the last and first columns of Tables~\ref{tab.swx1} and \ref{tab.swx2}, respectively,  with those determined by fitting directly the experimental data in Table~\ref{tab.210804.1}.

\section{Conclusions}
\label{sec.conc}

In this work we focus on the narrow peak around 4459~MeV observed in the $\jpsi\Lambda$ event distributions from the LHCb measurements~\cite{Aaij:2020gdg}, which is the first discovered charmonium pentaquark with strangeness. We have considered three different methods, based on the elastic effective-range expansion, the saturation of the width and the compositeness of the resonance, and a fit to data employing unitarized phenomenological amplitudes with constraints from heavy-quark-spin symmetry.

It is important to notice that the different methods provide compatible results for those magnitudes to which they can be applied for their calculation. In this sense, our analyses clearly favor a dominant role of the $\Xi_c\bar{D}^*$ state in the composition of the $P_{cs}(4459)$. For instance, the elastic effective-range analysis gives an effective range of $-0.94\pm 0.13$~fm when considering only the $\Xi_c\bar{D}^*$, while it is much larger in absolute value, around $-4~$fm, when taking only the $\Xi_c'\bar{D}$ channel. When considering the fits to data we then obtain that the compositeness factor $X$ of the $\Xi_c \bar{D}^*$ channel is around one, implying that it is saturated almost entirely by the $\Xi_c\bar{D}^*$ channel alone. Finally, even if allowing the presence of a CDD pole in the two-channel case, with $J/\psi\Lambda$ and either $\Xi_c\bar{D}^*$ or $\Xi_c'\bar{D}$, no acceptable fits result, which is also a clear indication of the molecular scenario.

 Another important outcome of our fits is that when requiring heavy-quark-spin symmetry, such that the $J=1/2$ and 3/2 couplings of the $\Xi_c\bar{D}^*$  are equal, our analysis requires to include also the $\Xi'_c\bar{D}$ channel to fit data properly. Furthermore, in such circumstances, the resulting fit contains two poles in the energy region of interest around the resonant peak of the $P_{cs}(4459)$, a  $J=1/2$ pole  with mass $4453.8^{+2.4}_{-3.3}$~MeV, and a $J=3/2$ one with mass $4466.6^{+1.9}_{-2.7}$~MeV. These two poles are narrow, the half-width obtained for the lighter one is  $2.8^{+0.9}_{-0.8}$~MeV, and that for the heavier one is determined much less precisely as $1.3^{+1.3}_{-3.7}$~MeV. With the present energy resolution and statistics of the LHCb collaboration \cite{Aaij:2020gdg} there is only a small hint that there are actually two resonances involved, though this could be disentangled in the future  by experiments with improved statistics.

\section*{Acknowledgements}
This work is funded in part by the Natural Science Foundation of China under Grant Nos.~11975090 and ~11575052, the MICINN AEI (Spain) Grant No. PID2019-106080GB-C22/AEI/10.13039/501100011033, and the Fundamental Research Funds for the Central Universities with contract No.~2242021R10099. The work of M.L.D. is supported by the Spanish Ministerio de Econom\'ia y Competitividad (MINECO) and the European Regional Development Fund (ERDF) under contract FIS2017-84038-C2-1-P, by the EU Horizon 2020 research and innovation programme, STRONG-2020 project, under grant agreement No.~824093, by Generalitat Valenciana under contract PROMETEO/2020/023. 

\bibliographystyle{ieeetr}
\bibliography{biblio.bib}

\end{document}